\documentclass{article}
\usepackage{spconf,amsmath,graphicx}

\usepackage{cite}
\usepackage{amsmath,amssymb,amsfonts, amsthm}
\usepackage{graphicx}
\usepackage{textcomp}
\usepackage{url}
\usepackage{mathbbol}
\usepackage{latexsym}
\usepackage{ascmac}
\usepackage{bm}
\usepackage{ifpdf}

\usepackage{subfigure}
\usepackage[ruled,vlined,linesnumbered,noresetcount]{algorithm2e}
\usepackage{multirow}
\usepackage{multicol}
\usepackage{array}
\usepackage{tabularx}
\usepackage{lineno,hyperref}

\theoremstyle{definition}
\newtheorem{dfn}{Definition}

\def\BibTeX{{\rm B\kern-.05em{\sc i\kern-.025em b}\kern-.08em
    T\kern-.1667em\lower.7ex\hbox{E}\kern-.125emX}}

\newcommand{\transp}{{\sf T}}

\newcommand{\argmin}{\mathop{\mathrm{arg\,min}}\limits}

\title{Graph Filter Transfer Via Probability Density Ratio Weighting}
\name{Koki Yamada$^{1}$ \thanks{This work was supported in part by KAKENHI under grant 22K21287.}}
\address{$^{1}$Department of Electrical Engineering, Tokyo University of Science, Tokyo, Japan}
\begin{document}
\ninept
\maketitle
\begin{abstract}
The problem of recovering graph signals is one of the main topics in graph signal processing.
A representative approach to this problem is the graph Wiener filter, which utilizes the statistical information of the target signal computed from historical data to construct an effective estimator. 
However, we often encounter situations where the current graph differs from that of historical data due to topology changes, leading to performance degradation of the estimator.
This paper proposes a graph filter transfer method, which learns an effective estimator from historical data under topology changes.
The proposed method leverages the probability density ratio of the current and historical observations and constructs an estimator that minimizes the reconstruction error in the current graph domain.
The experiment on synthetic data demonstrates that the proposed method outperforms other methods.
\end{abstract}
\section{Introduction}
\label{sec:intro}
Signals on the network appear in many fields, e.g., sensor networks, wireless communication, electrical power system, transportation network, etc. 
One of the popular tools for analyzing such data is graph signal processing (GSP) \cite{ortega2018, shuman2013}.
GSP enables frequency analysis, filtering, and sampling even for data with complex structures that cannot be handled by traditional signal processing.
GSP uses graphs to represent the network structure and provides efficient analysis taking into account the pairwise relationships between signals.

The main topics of GSP include graph signal recovery problems \cite{nagahama2022, isufi2017a, kroizer2022, routtenberg2021}. 
This problem aims to recover the desired signals from noisy and corrupted observations. 
The graph signal recovery problem arises in many applications: power system state estimation, missing data interpolation in sensor networks, image restoration, and point cloud processing \cite{jablonski2017, ramakrishna2021, ono2015}. 
The solutions to graph signal recovery are roughly categorized into deterministic and statistical approaches. 
The deterministic approaches generally formulate the problem as a convex problem based on the observation process and utilize regularizations that reflect the prior information on the signals to estimate the desired signals \cite{ono2015, romero2017}. 
The statistical approaches use the statistical information (mean and covariance) of the desired signals to construct an efficient estimator \cite{routtenberg2021, kroizer2022}.
The representative method of these is the graph Wiener filter, which is a natural extension of the linear minimum mean square error (LMMSE) estimator \cite{perraudin2017, marques2017, hara2020}. 
The graph Wiener filter assumes the stationarity of the graph signals, which means that the mean and covariance of them are unchanged with the graph shift.
The statistical information based on stationarity often yields a better average performance compared to deterministic approaches. 

Since statistical information on signals is not given in many practical applications, it is estimated from historical data. 
For example, the covariance in the graph Wiener filter is computed by the power spectral density (PSD) estimation in the graph domain \cite{marques2017}.
The PSD estimation assumes implicitly that the graph in historical data and the current graph are the same. 
However, this assumption often does not hold, resulting in situations where the historical and current graphs are different.
A typical example is seen in the power system, where the topology changes by switching the circuit breaker and adding/removing loads and generators \cite{grotas2019, srikantha2019}. 

The key research question addressed in this paper is: \emph{how should we construct an efficient estimator under a situation where the current topology differs from that of historical data?}
This problem is called \emph{graph filter transfer} hereafter.
A simple solution to this problem is to learn a parametric graph filter from historical data and transfer its parameters to the current graph domain, as proposed in \cite{kroizer2022}.
However, the estimator constructed in this way is only optimal for the graph in historical data, but not for the current graph domain.

This paper proposes a graph filter transfer method using the probability density ratio weighting.
The proposed method is inspired by transfer learning in machine learning \cite{pan2010}.
The strategy of the proposed method is to compute the importance of data samples in historical data based on the probability density ratio of current and historical observations and to learn a parametric graph filter from historical data weighted by its importance. 
This enables the construction of an estimator that minimizes the mean square error of the reconstruction in the current graph domain.
The experiment of missing-data interpolation on simulation data demonstrates that the estimator constructed by the proposed method successfully recovers graph signals. 

The remainder of this paper is organized as follows. 
We summarize the basics of GSP and the graph Wiener filter in Section 2. 
Section 3 presents the graph filter transfer method based on the probability density ratio weighting. 
Experimental results on simulation data are provided in Section 4.  
Conclusions are given in Section 5.

\noindent
\textit{Notation and Definitions:}
Lowercase normal, lowercase bold, and uppercase bold letters denote scalars, vectors, and matrices, respectively.
Calligraphic capital letters denote sets and the complement of a set $\mathcal{A}$ is denoted by $\mathcal{A}^{c}$ .
$\mathbf{x}_{\mathcal{S}}$ and $[\mathbf{x}]_{\mathcal{S}}$ represent the vector whose elements of indexes included in set $\mathcal{S}^{c}$ are eliminated.
$\mathbf{X}_{\mathcal{S}}$ and $[\mathbf{X}]_{\mathcal{S}}$ are the submatrix obtained by removing columns included in $\mathcal{S}^{c}$ from $\mathbf{X}$.
$\mathrm{N}(\boldsymbol{\mu}, \boldsymbol{\Sigma})$ is a multivariate Gaussian distribution with the mean $\boldsymbol{\mu}$ and the covariance $\boldsymbol{\Sigma}$.
The uniform distribution in the interval $[x, y]$ is denoted by $\mathrm{U}(x, y)$.

\section{Graph Wiener Filter}
\label{sec:gwfilter}
\subsection{Basic Definitions for GSP}
An undirected weighted graph is defined as $\mathcal{G} = (\mathcal{V}, \mathcal{E}, {\bf W})$, where $\mathcal{V}$ is a set of nodes, $\mathcal{E}$ is a set of edges, and $\mathbf{W}$ is a weighted adjacency matrix. 
$N =|\mathcal{V}|$ represents the number of nodes.
The degree matrix ${\bf D}$ is a diagonal matrix whose diagonal element is $d_{mm} = \sum_n w_{mn}$.
The graph Laplacian is given by ${\bf L} = {\bf D} - {\bf W}$.
Since $\bf L$ is a real symmetric matrix, it has orthogonal eigenvectors and can be decomposed into ${\bf L} = {\bf U}{\bf \Lambda} {\bf U}^{\transp}$, where ${\bf U} = [{\bf u}_{0}, {\bf u}_{1}, \ldots , {\bf u}_{N-1}]$ is a matrix whose $i$-th column is the eigenvector ${\bf u}_{i}$ and ${\bf \Lambda} = {\rm diag}(\lambda_{0}, \lambda_{1}, \ldots \lambda_{N-1})$ is a diagonal eigenvalue matrix.
The eigenvalues $\lambda_{i}$ are sorted in ascending order.

The graph Fourier transform (GFT) and the inverse GFT are defined as $\hat{\mathbf{x}} = \mathbf{U}^{\transp}\mathbf{x}$ and $\mathbf{x} = \mathbf{U}\hat{\mathbf{x}}$.
The eigenvalues of the Laplacian graph correspond to the frequencies in the classical Fourier transform, and thus are often called \emph{graph frequencies}.

\subsection{Graph Wide-Sense Stationary and LMMSE Estimator}
Graph stationary plays an essential role in statistical graph signal processing.
First, we introduce the definition of a graph wide-sense stationary.
\begin{dfn}[Graph wide-sense stationary \cite{perraudin2017}]
A stochastic graph signal defined on a graph $\mathcal{G}$ is called Graph Wide-Sense Stationary (GWSS), if and only if the following conditions are satisfied:
\begin{enumerate}
    \item $\mathbb{E}[\mathbf{x}] = \boldsymbol{\mu}_{\mathbf{x}} = \mathrm{const.}$
    \item $\boldsymbol{\Sigma}_{x} =\mathbf{U} \mathrm{diag}({\mathbf{p}}) \mathbf{U}^{\transp}$
where $\mathbf{p}$ is called the power spectral density (PSD).
\end{enumerate}
\end{dfn}
GWSS is an extension of wide-sense stationary in the time domain and has the property that the covariance of stochastic graph signals can be diagonalized by the GFT matrix $\mathbf{U}$.
This property is the counterpart of the Wiener–Khinchin theorem in traditional signal processing.
The study in \cite{perraudin2017} demonstrates that graph signals with GWSS appear in many applications.

We consider the problem of recovering stochastic graph signals $\mathbf{x} \in \mathbb{R}^{N}$ on $\mathcal{G}$ from observations $\mathbf{y} \in \mathbb{R}^{d}$
based on the following observation model:
\begin{equation}
    \mathbf{y} = \mathbf{Mx} + \epsilon, \ \mathbf{x} \sim q^{*}(\mathbf{x} | \mathbf{L}),
    \label{eq:observation}
\end{equation}
where $\mathbf{M}$ is a degradation matrix,  $q^{*}(\mathbf{x} | \mathbf{L})$ is a conditional distribution of $\mathbf{x}$ given a graph Laplacian $\mathbf{L}$ with the mean $\boldsymbol{\mu}_{\mathbf{x}}$ and the covariance $\boldsymbol{\Sigma}_{\mathbf{x}}$, and $\epsilon \sim \mathcal{N}(0, \sigma\mathbf{I})$ is an additive white Gaussian noise.
In this problem setting, we assume that $\mathbf{M}$, $\sigma$ and $\mathbf{L}$ are given.
From \eqref{eq:observation} we find a linear minimum mean square error (LMMSE) estimator by solving the following problem \cite{kay1993}:
\begin{equation}
    \{ \hat{\mathbf{Q}}, \hat{\mathbf{b}} \}=\argmin_{\mathbf{Q}, \mathbf{b}} \mathbb{E}\left[ \| \mathbf{Qy} + \mathbf{b} - \mathbf{x} \|_{2}^{2} \right].
    \label{eq:lmmse}
\end{equation}
Since \eqref{eq:lmmse} has closed form solution, $\hat{\mathbf{Q}}$ and $\hat{\mathbf{b}}$ are given by:
\begin{align}
    \hat{\mathbf{Q}} &= \boldsymbol{\Sigma}_{\mathbf{x}} \mathbf{M}^{\transp}(\mathbf{M}\boldsymbol{\Sigma}_{\mathbf{x}}  \mathbf{M}^{\transp})^{-1}, \label{eq:gwf_q} \\
    \hat{\mathbf{b}} &= \boldsymbol{\mu}_{x} - \hat{\mathbf{Q}} \mathbb{E}[ \mathbf{\mathbf{y}} ].
\end{align}
When $\mathbf{x}$ satisfies the GWSS conditions, i.e., $\Sigma_{\mathbf{x}} = \mathbf{U} \mathrm{diag}({\mathbf{p}}) \mathbf{U}^{\transp}$ this LMMSE estimator is called \emph{graph Wiener filter} \cite{perraudin2017, hara2020}.

\subsection{Power Spectral Density Estimator}
In many applications, the mean $\boldsymbol{\mu}_{x}$ and covariance $\boldsymbol{\Sigma}_{\mathbf{x}}$ in \eqref{eq:observation} are unknown, and thus it is required to estimate them from historical data (training data) ${\mathbf{x}^{(1)}, \ldots \mathbf{x}^{(K)}}$.
Since the mean of $\mathbf{x}$ can be easily obtained by computing the sample mean, we focus on the covariance estimation here.

A simple solution of the covariance estimation is to compute the sample covariance; however, it requires a lot of data samples and the estimated covariance is not diagonalizable with GFT matrix $\mathbf{U}$ in many cases. 
The covariance estimation under the GWSS assumption is formulated as the estimation of the PSD of the graph signal.
The PSD can be represented as the expectation of the squared GFT of the graph signal, that is, a nonparametric PSD estimator is given by \cite{marques2017}:
\begin{equation}
    \hat{\mathbf{p}}(\mathbf{x}) = \frac{1}{K} \sum_{k= 1}^{K} (\hat{\mathbf{x}}^{(k)})^{2},
\end{equation}
where $\hat{\mathbf{x}}^{(k)} = \mathbf{U}^{\transp} \mathbf{x}^{(k)}$.

Next, we introduce a parametric PSD estimation, which approximates the PSD by the parametric graph filter.
This approach finds parameters by solving the following optimization problem:
\begin{equation}
    \hat{\boldsymbol{\beta}} = \argmin_{\boldsymbol{\beta}} D_{\mathbf{p}} ( \hat{\mathbf{p}}(\mathbf{x}), |\mathbf{f}(\boldsymbol{\beta})|^{2}),
    \label{eq:psde_sq}
\end{equation}
where $\mathbf{f}(\boldsymbol{\beta})$ is a graph filter with parameter $\beta$ to fit the PSD, and $D_{p}$ is a metric to compare $\hat{\mathbf{p}}(\mathbf{x})$ and $|\mathbf{f}(\boldsymbol{\beta})|^{2}$.
Although the metric is often set to the square $\ell_{2}$-norm, $D_{\mathbf{p}} (  \hat{\mathbf{p}}(\mathbf{x}), |\mathbf{f}(\boldsymbol{\beta})|^{2}) = \| | \hat{\mathbf{p}}(\mathbf{x}) - |\mathbf{f}(\boldsymbol{\beta})|^{2}  \|_{2}^{2} $, \eqref{eq:psde_sq} becomes a nonconvex and intractable problem.
If $\mathbf{\beta}$ is constrained such that $\mathbf{p}(\boldsymbol{\beta})$ is nonnegtive, \eqref{eq:psde_sq} can be reduce to the problem given by:
\begin{equation}
    \hat{\boldsymbol{\beta}} = \argmin_{\boldsymbol{\beta} \in \Omega_{\boldsymbol{\beta}}} D_{\mathbf{p}} ( \sqrt{\hat{\mathbf{p}}(\mathbf{x})}, \mathbf{f}(\boldsymbol{\beta})),
    \label{eq:psde}
\end{equation}
where $\Omega_{\boldsymbol{\beta}}$ is the relevant parameter space.
Since this problem is a convex problem with nonnegative constraints, it can be solved using a convex optimization algorithm \cite{boyd2004}.

\section{Graph Filter Transfer}
\label{sec:glearn}

\subsection{Problem Formulation}
We consider the graph filter transfer problem, which designs an effective estimator under the situation that the current graph is differ from that of historical data.
Let $\mathcal{G}_{\mathrm{h}}$ and $\mathcal{G}_{\mathrm{c}}$ be the historical and current graphs, respectively.
For simplicity, we assume that $\mathcal{G}_{\mathrm{h}}$ and $\mathcal{G}_{\mathrm{c}}$ have the same number of nodes, $N_{\mathrm{h}} = N_{\mathrm{c}}$, and the case of $N_{\mathrm{h}} \neq N_{\mathrm{c}}$ is handled in Section \ref{sbsec:nodechang}.
The observation model of the graph filter transfer problem can be represented as follows:
\begin{align}
    &\mathbf{y}_{\mathrm{h}} = \mathbf{M}\mathbf{x}_{\mathrm{h}} + \epsilon, \ \mathbf{x}_h \sim q^{*}_{\mathrm{h}}(\mathbf{x} | \mathbf{L}_{\mathrm{h}}), \label{eq:obs_h} \\
    &\mathbf{y}_{\mathrm{c}} = \mathbf{M}\mathbf{x}_{\mathrm{c}} + \epsilon, \ \mathbf{x}_c \sim q^{*}_{\mathrm{c}}(\mathbf{x} | \mathbf{L}_{\mathrm{c}}).
    \label{eq:obs_c}
\end{align}
Note that \eqref{eq:obs_h} and \eqref{eq:obs_c} use the same notation as \eqref{eq:observation}, with the subscript h for historical data and the subscript c for current data.
In this problem setting, the historical data $\{\mathbf{x}_{\mathrm{h}}^{(1)} \mathbf{y}_{\mathrm{h}}^{(1)}\}, \cdots \{\mathbf{x}_{\mathrm{h}}^{(K_{\mathrm{h}})},  \mathbf{y}_{\mathrm{h}}^{(K_{\mathrm{h}})}\}$ and the observations $\mathbf{y}_{\mathrm{c}}^{(1)}, \cdots \mathbf{y}_{\mathrm{c}}^{(K_{\mathrm{c}})}$ are given, but no dataset for $\mathbf{x}_{\mathrm{c}}$ is provided.
Assume that $\mathbf{x}_{\mathbf{h}}$ and $\mathbf{x}_{\mathbf{c}}$ are GWSS, and their covariances are given by
\begin{equation}
    \Sigma_{\mathbf{x}_{\mathrm{h}}} = \mathbf{U}_{\mathrm{h}} \mathrm{diag}(\mathbf{p}_{\mathrm{h}}) \mathbf{U}_{\mathrm{h}}^{\transp}, \ \Sigma_{\mathbf{x}_{\mathrm{c}}} = \mathbf{U}_{\mathrm{c}} \mathrm{diag}(\mathbf{p}_{\mathrm{c}}) \mathbf{U}_{\mathrm{c}}^{\transp}.
\end{equation}

A simple solution to this problem is to estimate $\mathbf{\beta}$ from historical data and use $\mathbf{\beta}$ to construct the estimator of $\mathbf{x}_{\mathrm{c}}$. 
Specifically, it first solves the following optimization problem,
\begin{equation}
    \hat{\boldsymbol{\beta}}_{\mathrm{h}} = \argmin_{\boldsymbol{\beta} \in \Omega_{\boldsymbol{\beta}}}  D_{\mathbf{p}}(  \sqrt{\hat{\mathbf{p}}(\mathbf{x}_{\mathrm{h}})}, \mathbf{f}(\boldsymbol{\beta})),
    \label{eq:simple_gftr}
\end{equation}
and construct the covariance: $\hat{\Sigma}_{\mathbf{x}_{\mathrm{c}}} = \mathbf{U}_{\mathrm{c}}\mathrm{diag}(\mathbf{f}(\hat{\boldsymbol{\beta}}_{\mathrm{h}})) \mathbf{U}_{\mathrm{c}}^{\transp}.$
Substituting $\hat{\Sigma}_{\mathbf{x}_{\mathrm{c}}}$ into \eqref{eq:gwf_q}, we can obtain the estimator of $\mathbf{x}_{\mathrm{c}}$.
The estimated $\hat{\boldsymbol{\beta}}_{\mathrm{h}}$ in this way is optimal only for historical data, but not for current data.

The proposed method aims at learning an optimal $\hat{\boldsymbol{\beta}}$ for the current data, i.e., our goal is to solve the following problem:
\begin{equation}
    \hat{\boldsymbol{\beta}} = \argmin_{\boldsymbol{\beta} \in \Omega_{\boldsymbol{\beta}}} \mathbb{E}_{\mathbf{x}_c \sim q^{*}_{\mathrm{c}}(\mathbf{x})}[D_{\mathbf{p}} (  \sqrt{\hat{\mathbf{p}}(\mathbf{x}_{\mathrm{c}})}, \mathbf{f}(\boldsymbol{\beta}))].
    \label{eq:dp_emp}
\end{equation}
Since the dataset of $\mathbf{x}_{\mathrm{c}}$ is not given and the expectation over $\mathbf{x}_{\mathrm{c}}$ cannot be computed, it seems impossible to solve \eqref{eq:dp_emp}.
Leveraging the technique of importance sampling\cite{sugiyama2012}, \eqref{eq:dp_emp} can be transformed into a tractable problem as follows:
\begin{equation}
\begin{split}
    & \mathbb{E}_{\mathbf{x}_c \sim q^{*}_{\mathrm{c}}(\mathbf{x})}[D_{\mathbf{p}} (  \sqrt{\hat{\mathbf{p}}(\mathbf{x})}, \mathbf{f}(\boldsymbol{\beta}))] \\
    & = \int D_{\mathbf{p}} (  \sqrt{\hat{\mathbf{p}}(\mathbf{x})}, \mathbf{f}(\boldsymbol{\beta})) q^{*}_{\mathrm{c}}(\mathbf{x} | \mathbf{L}_{\mathrm{c}}) \mathrm{d}\mathbf{x} \\
    & = \int D_{\mathbf{p}} (  \sqrt{\hat{\mathbf{p}}(\mathbf{x})}, \mathbf{f}(\boldsymbol{\beta})) \frac{q^{*}_{\mathrm{c}}(\mathbf{x} | \mathbf{L}_{\mathrm{c}})}{q^{*}_{\mathrm{h}}(\mathbf{x} | \mathbf{L}_{\mathrm{h}})} q^{*}_{\mathrm{h}}(\mathbf{x} | \mathbf{L}_{\mathrm{h}} ) \mathrm{d}\mathbf{x} \\
    & = \mathbb{E}_{\mathbf{x}_{\mathrm{h}} \sim q^{*}_{\mathrm{h}}(\mathbf{x})} \left[D_{\mathbf{p}} (  \sqrt{\hat{\mathbf{p}}(\mathbf{x})}, \mathbf{f}(\boldsymbol{\beta})) \frac{q^{*}_{\mathrm{c}}(\mathbf{x}_{\mathrm{h}} | \mathbf{L}_{\mathrm{c}})}{q^{*}_{\mathrm{h}}(\mathbf{x}_{\mathrm{h}} | \mathbf{L}_{\mathrm{h}})} \right].
    \label{eq:dp_emp_trs}
\end{split}
\end{equation}
By Bayes' theorem, the density ratio $ \frac{q^{*}_{\mathrm{c}}(\mathbf{x} | \mathbf{L}_{\mathrm{c}})}{q^{*}_{\mathrm{h}}(\mathbf{x} | \mathbf{L}_{\mathrm{h}})}$ in \eqref{eq:dp_emp_trs} can be transformed into the following form:
\begin{equation}
    \frac{q^{*}_{\mathrm{c}}(\mathbf{x} | \mathbf{L}_{\mathrm{c}})}{q^{*}_{\mathrm{h}}(\mathbf{x} | \mathbf{L}_{\mathrm{h}})}
    = \frac{q^{*}_{\mathrm{c}}(\mathbf{y} | \mathbf{L}_{\mathrm{c}}) q(\mathbf{x} | \mathbf{y}) / q(\mathbf{y} | \mathbf{x}) }{q^{*}_{\mathrm{h}}(\mathbf{y} | \mathbf{L}_{\mathrm{h}}) q(\mathbf{x} | \mathbf{y}) / q(\mathbf{y} | \mathbf{x})}
    = \frac{q^{*}_{\mathrm{c}}(\mathbf{y} | \mathbf{L}_{\mathrm{c}})}{q^{*}_{\mathrm{h}}(\mathbf{y} | \mathbf{L}_{\mathrm{h}})} = r^{*}(\mathbf{y}).
    \label{eq:dratio_y}
\end{equation}
Substituting \eqref{eq:dratio_y} into \eqref{eq:dp_emp_trs}, we can obtain the following equation:
\begin{equation}
\begin{split}
    &\mathbb{E}_{\mathbf{x}_c \sim q^{*}_{\mathrm{c}}(\mathbf{x})}[D_{\mathbf{p}} (  \sqrt{\hat{\mathbf{p}}(\mathbf{x})}, \mathbf{f}(\boldsymbol{\beta}))] \\ 
    &= \mathbb{E}_{\mathbf{x}_{\mathrm{h}}, \mathbf{y}_{\mathrm{h}} \sim q^{*}_{\mathrm{h}}(\mathbf{x}, \mathbf{y})} \left[D_{\mathbf{p}} (  \sqrt{\hat{\mathbf{p}}(\mathbf{x})}, \mathbf{f}(\boldsymbol{\beta})) r^{*}(\mathbf{y}) \right].
    \label{eq:dp_emp_f}
\end{split}
\end{equation}
This implies that if the density ratio $r^{*}(\mathbf{y})$ is obtained, \eqref{eq:dp_emp} can be solved by minimizing the empirical average of \eqref{eq:dp_emp_f} over historical data $\{ \mathbf{x}_{\mathrm{h}}, \mathbf{y}_{\mathrm{h}}\}$:
\begin{equation}
    \argmin_{\boldsymbol{\beta} \in \Omega_{\boldsymbol{\beta}}} D_{\mathbf{p}} ( \sqrt{\hat{\mathbf{p}}_{\mathrm{r}}(\mathbf{x})}, \mathbf{f}(\boldsymbol{\beta})),
\label{eq:opt_gft_drw}
\end{equation}
where
\begin{equation}
    \hat{\mathbf{p}}_{\mathrm{r}}(\mathbf{x}) = \frac{1}{K_{\mathrm{h}}} \sum_{k= 1}^{K_{\mathrm{h}}} r^{*}(\mathbf{y}_{\mathrm{h}}^{(k)}) (\mathbf{U}_{\mathrm{c}}\mathbf{x}_{\mathrm{h}}^{(k)})^{2}.
\end{equation}
The computation of $r^{*}(\mathbf{y})$ is presented in Section \ref{sec:cpdr}.

\subsection{Computation of Probability Density Ratio}
\label{sec:cpdr}
The density ratio $r^{*}(\mathbf{y})$ estimation can be formulated as the problem of learning the model $r(\mathbf{y})$ that minimizes the following squared error $\mathbf{J}$ \cite{kanamori2009}:
\begin{equation}
\begin{split}
    J &= \frac{1}{2} \int \left( r(\mathbf{y}) - r^{*}(\mathbf{y}) q^{*}_{\mathrm{h}}(\mathbf{y} | \mathbf{L}_{\mathrm{h}}) \right)^{2} \mathrm{d}\mathbf{y} \\
    &= \frac{1}{2} \int  r(\mathbf{y})^{2} q^{*}_{\mathrm{h}}(\mathbf{y} | \mathbf{L}_{\mathrm{h}})  \mathrm{d}\mathbf{y}
    - \int  r(\mathbf{y}) q^{*}_{\mathrm{c}}(\mathbf{y} | \mathbf{L}_{\mathrm{c}}) \mathrm{d}\mathbf{y} \\
    &+ \frac{1}{2} \int  r^{*}(\mathbf{y}) q^{*}_{\mathrm{c}}(\mathbf{y} | \mathbf{L}_{\mathrm{c}})  \mathrm{d}\mathbf{y}.
\end{split}
\end{equation}
Approximating the expectations in $J$ by empirical averages, the optimization problem is given by:
\begin{equation}
\begin{split}
    &\argmin_{r} J \\
    &= \argmin_{r} \frac{1}{2} \int  r(\mathbf{y})^{2} q^{*}_{\mathrm{h}}(\mathbf{y} | \mathbf{L}_{\mathrm{h}})  \mathrm{d}\mathbf{y}
    - \int  r(\mathbf{y}) q^{*}_{\mathrm{c}}(\mathbf{y} | \mathbf{L}_{\mathrm{c}}) \mathrm{d}\mathbf{y} \\
    &\simeq \argmin_{r} \frac{1}{2K_{\mathrm{h}}} \sum_{k=1}^{K_h}r\left(\mathbf{y}_{\mathrm{h}}^{(k)}\right)^{2} - \frac{1}{K_{\mathrm{c}}} \sum_{k=1}^{K_{\mathrm{c}}} r\left(\mathbf{y}_{\mathrm{c}}^{(k)}\right).
    \label{eq:opt_dre}
\end{split}
\end{equation}
We define the model $r(\mathbf{y})$ as follow:
\begin{equation}
    r(\mathbf{y}) = \sum_{l=1}^{b} \theta_{l}\phi_{l}(\mathbf{y}) = \boldsymbol{\phi}(\mathbf{y})^{\transp}\boldsymbol{\theta},
    \label{eq:r_model}
\end{equation}
where $\boldsymbol{\phi}(\mathbf{y}) : \mathbb{R}^{d} \rightarrow \mathbb{R}^{b}$ is a nonnegative basis function and $\boldsymbol{\theta}$ is a parameter.
Substituting \eqref{eq:r_model} into \eqref{eq:opt_dre} and adding the Tikhonov regularization, we obtain the following problem: 
\begin{equation}
    \argmin_{\boldsymbol{\theta}} \frac{1}{2} \boldsymbol{\theta}^{\transp}\mathbf{H}\boldsymbol{\theta}
    - \mathbf{h}^{\transp}\boldsymbol{\theta} + \frac{\lambda}{2} \boldsymbol{\theta}^{\transp}\boldsymbol{\theta},
    \label{eq:dr_opt}
\end{equation}
where
\begin{equation}
    \mathbf{H} = \frac{1}{K_{\mathrm{h}}}\sum_{k=1}^{K_{\mathrm{h}}} \boldsymbol{\phi}(\mathbf{y}_{\mathrm{h}}^{(k)})\boldsymbol{\phi}(\mathbf{y}_{\mathrm{h}}^{(k)})^{\transp}, \ 
    \mathbf{h} = \frac{1}{K_{\mathrm{c}}}\boldsymbol{\phi}(\mathbf{y}_{\mathrm{c}}^{(k)}).
    \label{eq:dr_Hh}
\end{equation}
This optimization problem can be solved efficiently using the algorithm in \cite{kanamori2009}.

\subsection{Learning ARMA graph filter}
Using the density ratio $r(\mathbf{y})$ estimated in Section \ref{sec:cpdr}, we solve \eqref{eq:opt_gft_drw} to learn a parametric graph filter. 
The parametric graph filter used in the proposed method is the autoregressive moving average (ARMA) graph filter given by \cite{isufi2017a, kroizer2022}:
\begin{equation}
    [\mathbf{f}_{\mathrm{ARMA}}(\boldsymbol{\alpha}, \boldsymbol{\beta})]_{i} = \frac{\sum_{l=0}^{L}\beta_{l} \lambda_{i}^{l}}{1 + \sum_{m=1}^{M} \alpha_{m}\lambda_{i}^{m}},
    \label{eq:armagf}
\end{equation}
where $\lambda_{i}$ is the eigenvalue of the graph Laplacian $\mathbf{L}_{\mathrm{c}}$.
Adopting the square $\ell_{2}$-norm for $D_{\mathbf{p}}$ and substituting \eqref{eq:armagf} to \eqref{eq:opt_gft_drw}, we obtain the following optimization problem:
\begin{equation}
    \argmin_{\boldsymbol{\alpha} \in \Omega_{\boldsymbol{\alpha}}, \boldsymbol{\beta} \in \Omega_{\boldsymbol{\beta}}} \| \sqrt{\hat{\mathbf{p}}_{\mathrm{r}}(\mathbf{x})} - (\mathrm{diag}(\mathbf{1} + \boldsymbol{\Phi}_{1}\boldsymbol{\alpha}))^{-1} \boldsymbol{\Phi}_{2}\boldsymbol{\beta} \|_{2}^{2},
    \label{eq:dp_arma_1}
\end{equation}
where
\begin{equation}
    \boldsymbol{\Phi}_{1} = \left[\begin{array}{ccc}\lambda_{1} & \ldots & \lambda_{1}^{M} \\ \vdots & \ddots & \vdots \\ \lambda_{N} & \ldots & \lambda_{N}^{M}\end{array}\right], \ 
    \boldsymbol{\Phi}_{2} = \left[\begin{array}{cccc}1 & \lambda_{1} & \ldots & \lambda_{1}^{L} \\ \vdots & \vdots & \ddots & \vdots \\ 1 & \lambda_{N} & \ldots & \lambda_{N}^{L}\end{array}\right].
    \label{eq:dp_alt}
\end{equation}
Since \eqref{eq:dp_arma_1} is an intractable problem, we replace it with the alternative one:
\begin{equation}
\begin{split}
    \argmin_{\boldsymbol{\alpha}, \boldsymbol{\beta}} & \| \mathbf{P}(\mathbf{1} + \boldsymbol{\Phi}_{1}\boldsymbol{\alpha}) - \boldsymbol{\Phi}_{2}\boldsymbol{\beta} \|_{2}^{2}
    + \boldsymbol{\alpha}^{\transp} \mathbf{R}_{\alpha} \boldsymbol{\alpha} + \boldsymbol{\beta}^{\transp} \mathbf{R}_{\beta} \boldsymbol{\beta}, \\
    &\mathrm{s.t.} \ \mathbf{1} + \boldsymbol{\Phi}_{1}\boldsymbol{\alpha} \geq \mathbf{0}, \ \boldsymbol{\Phi}_{2}\boldsymbol{\beta} \geq \mathbf{0},
\end{split}
\label{eq:opt_last}
\end{equation}
where $\mathbf{P} = \mathrm{diag}(\sqrt{\hat{\mathbf{p}}_{\mathrm{r}}(\mathbf{x})})$, $\boldsymbol{\alpha}^{\transp} \mathbf{R}_{\alpha} \boldsymbol{\alpha}$ and $\boldsymbol{\beta}^{\transp} \mathbf{R}_{\beta}\boldsymbol{\beta}$ are the regularization terms, and $\mathbf{R}_{\alpha}$ and $\mathbf{R}_{\beta}$ are positive semideﬁnite regularization matrices.
Since \eqref{eq:opt_last} is the convex optimization problem with the nonnegative constraint, it can be solved by the convex optimization algorithm, e.g., the augmented Lagrangian method \cite{komodakis2015}.

\subsection{Estimation Under Node Changes}
\label{sbsec:nodechang}
We consider the situation that $\mathcal{G}_h$ and $\mathcal{G}_{\mathrm{c}}$ have the different number of nodes, $N_{\mathrm{h}} \neq N_{\mathrm{c}}$. 
The sets of added and removed nodes are denoted by $\mathcal{V}_{a}$ and $\mathcal{V}_{r}$, that is, $N_{\mathrm{c}} = N_{\mathrm{h}} + |\mathcal{V}_{a} | - |\mathcal{V}_{r}|$.

The proposed method consists of the estimation of $r(\mathbf{y})$ and the learning of a parametric graph filter.
Let $\bar{\mathcal{V}} = \mathcal{V}_{\mathrm{h}} \cap \mathcal{V}_{\mathrm{c}}$, and we can compute $r(\mathbf{y})$ from \eqref{eq:dr_opt} and \eqref{eq:dr_Hh} using $[\mathbf{y}_{\mathbf{h}}]_{\bar{\mathcal{V}}}$ and $[\mathbf{y}_{\mathbf{c}}]_{\bar{\mathcal{V}}}$.
In the learning graph filter step, we first compute $\hat{\mathbf{p}}_{\mathrm{r}}(\mathbf{x})$ as follows:
\begin{equation}
    \hat{\mathbf{p}}_{\mathrm{r}}(\mathbf{x}) = \frac{1}{K_{\mathrm{h}}} \sum_{k=1}^{K_{\mathrm{h}}} r([\mathbf{y}_{\mathrm{h}}^{(k)}]_{\bar{\mathcal{V}}}) \left([\mathbf{U}_{\mathrm{c}}^{\transp}]_{\mathcal{V}_{a}^{c}} [\mathbf{I}_{\mathcal{V}_{r}^{c}}]^{\transp} \mathbf{x}_{\mathrm{h}}^{(k)}\right)^{2}.
    \label{eq:p_nodechange}
\end{equation}
Substitute \eqref{eq:p_nodechange} into \eqref{eq:opt_last}, we can learn the parametric graph filter under the situation that the number of nodes changes.

\section{Experiments}
\label{sec:exp_tm}

\begin{figure}[tb]
    \centering
    \subfigure[Ground truth]{\includegraphics[width=0.48\linewidth]{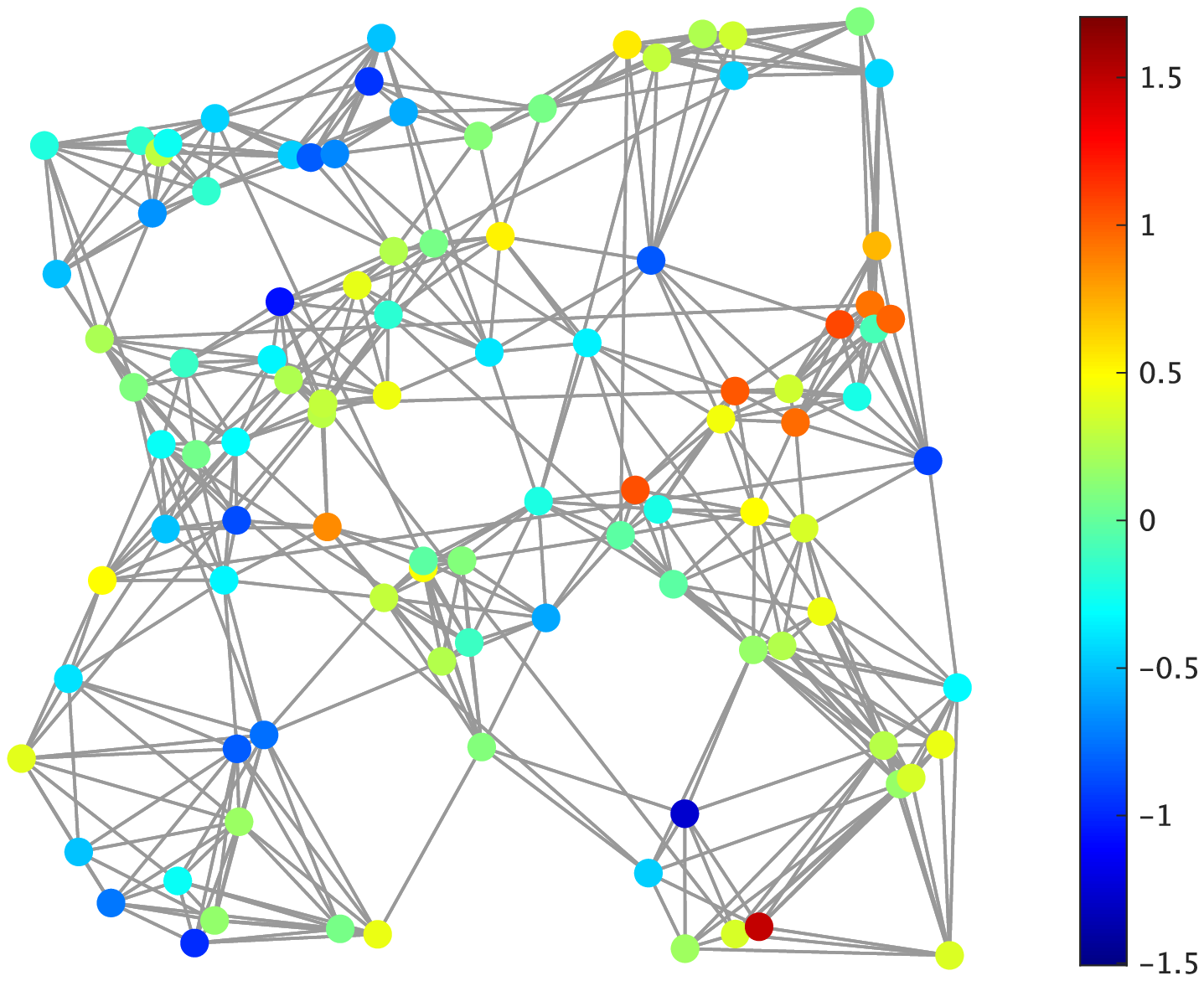}}
    \subfigure[Noisy data. MSE $ = 1.43 \times 10^{-1}$]{\includegraphics[width=0.48\linewidth]{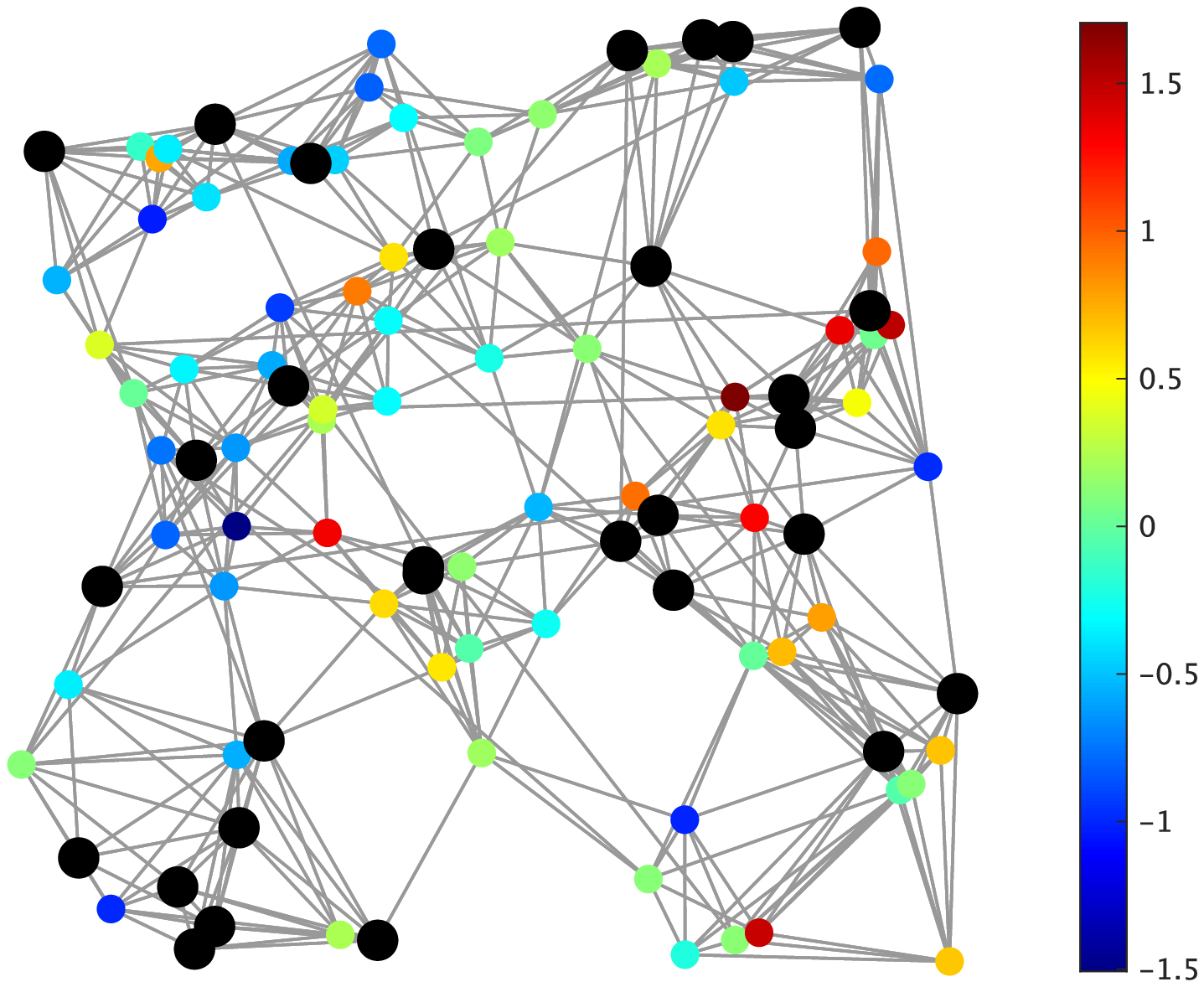}} \\
    \subfigure[ARMAE. MSE $ = 1.23 \times 10^{-1}$]{\includegraphics[width=0.48\linewidth]{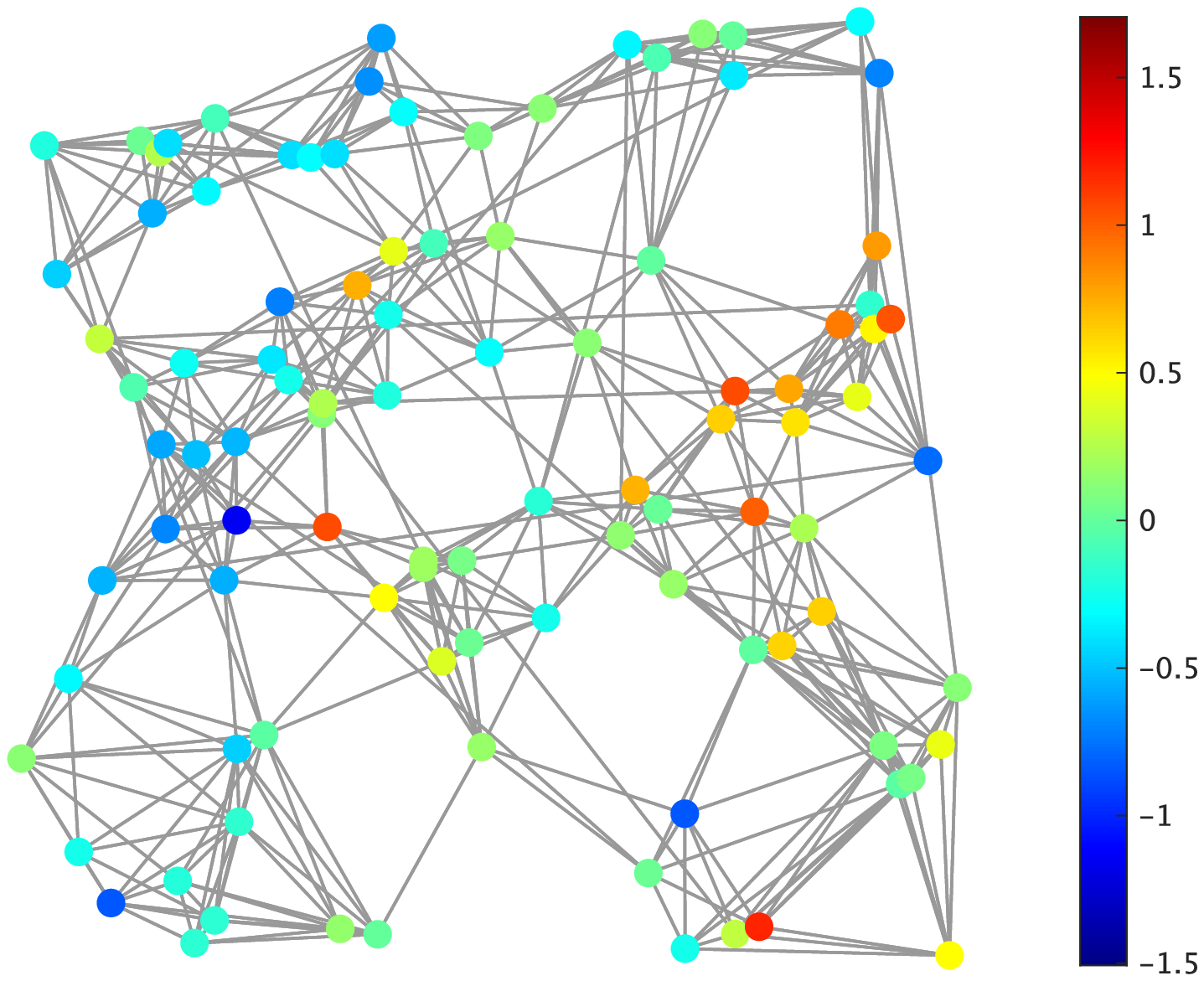}}
    \subfigure[ARMAE-DRW. MSE $ = 1.04 \times 10^{-1}$]{\includegraphics[width=0.48\linewidth]{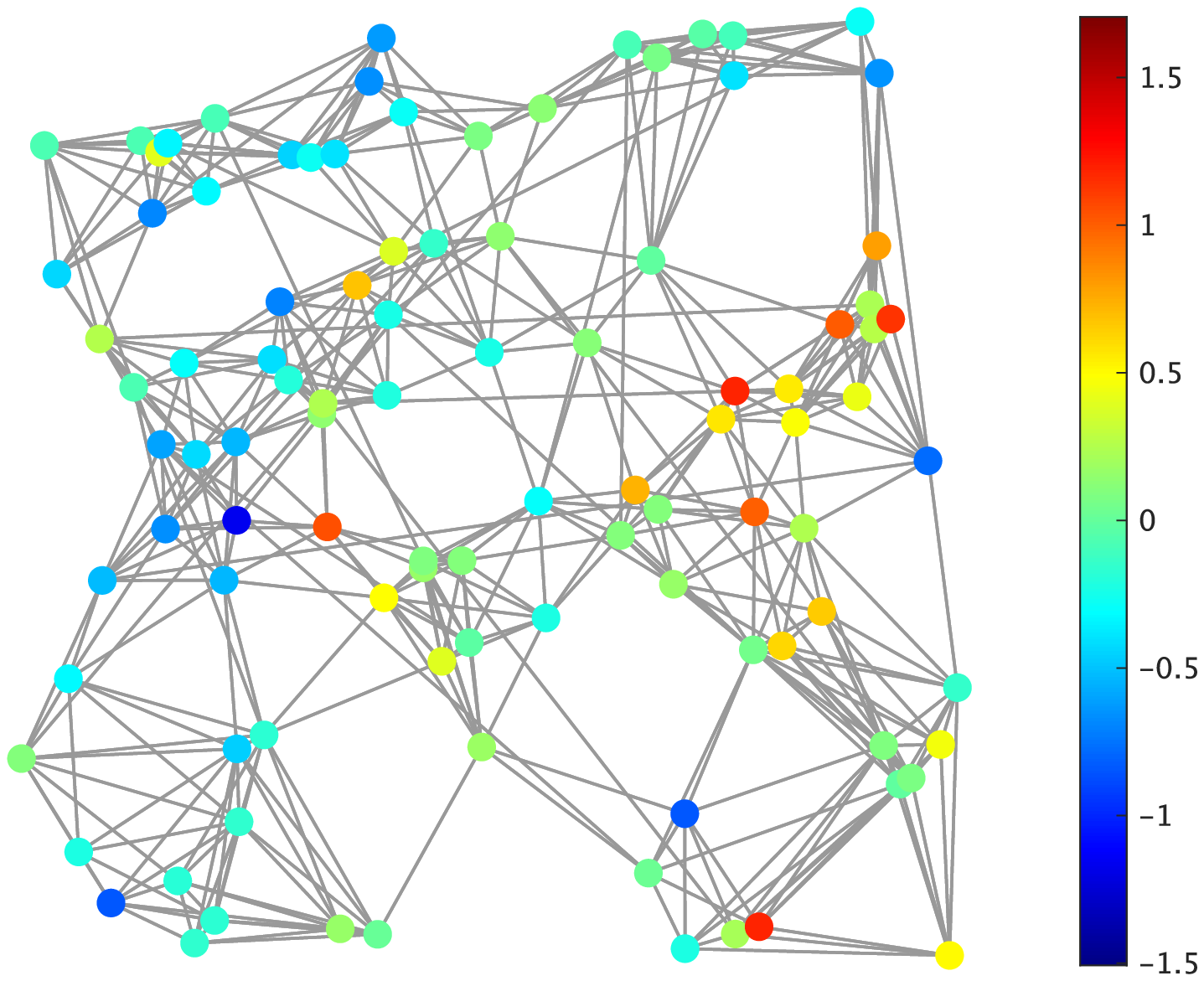}}
    \caption{The visualizations of the graph signal recovery in the RS graph. The black nodes in (b) represent the nodes for missing signals.}
    \label{fig:vis_gsr}
\end{figure}

\begin{table}[tb]
	\centering
	\caption{Average MSE of the recovered signals under edge changes.}
	\label{tab:echange}
	\begin{tabular}{c | wc{1cm} | wc{1.8cm} | wc{1.8cm}} \hline \hline
		\multirow{2}{*}{Methods} & \multirow{2}{*}{$e$} & \multicolumn{2}{c}{MSE ($\times 10^{-2})$} 
		\\ \cline{3-4} & & ER graph &  RS graph  \\ \hline \hline
		\multirow{3}{*}{ARMAE} 
		& 10  & 7.23 & 14.21 \\
		& 20  & 7.39 & 13.76 \\
	    & 30  & 7.13 & 14.41 \\ \hline \hline
	    \multirow{3}{*}{ARMAE-DRW} 
	    & 10  & \textbf{7.08} & \textbf{14.00} \\
		& 20  & \textbf{7.24} & \textbf{13.57} \\
	    & 30  & \textbf{6.98} & \textbf{13.81} \\  \hline \hline
	\end{tabular}
\end{table}

\begin{table}[tb]
	\centering
	\caption{Average MSE of the recovered signals under node changes.}
	\label{tab:nchange}
	\begin{tabular}{c | wc{1cm} | wc{1.8cm} | wc{1.8cm}} \hline \hline
		\multirow{2}{*}{Methods} & \multirow{2}{*}{$v$} & \multicolumn{2}{c}{MSE ($\times 10^{-2})$} 
		\\ \cline{3-4} & & ER graph &  RS graph  \\ \hline \hline
		\multirow{3}{*}{ARMAE} 
		& 10  & 7.33 & 13.53 \\
		& 20  & 7.03 & 14.29 \\
	    & 30  & 6.98 & 10.30 \\ \hline \hline
	    \multirow{3}{*}{ARMAE-DRW} 
	    & 10  & \textbf{6.97} & \textbf{11.61} \\
		& 20  & \textbf{6.62} & \textbf{10.00} \\
	    & 30  & \textbf{6.27} & \textbf{7.86 }\\  \hline \hline
	\end{tabular}
\end{table}

In this section, we present the experimental results of the graph signal recovery from noisy and missing data. 
We assume that the nodes for missing signals are known. 
The degradation matrix is given by $\mathbf{M} = \mathbf{I}_{\mathcal{S}}^{\transp}$, where $\mathcal{S}$ is the set of nodes for non-missing signals.
The missing probability is set to $30$\%.
The additive white Gaussian noise $\epsilon$ is generated from $\mathrm{N}(0, 0.1)$.

The power spectral densities $\mathbf{p}_{\mathrm{h}}$ and $\mathbf{p}_{\mathrm{c}}$ used for the experiment are given by $[\mathbf{p}_{\mathrm{h}}]_{i} = 1 - \lambda_{i} / \lambda_{\mathrm{max}}$ and $[\mathbf{p}_{\mathrm{c}}]_{i} = 1 / \lambda_{i}$.
We generate $2,000$ historical data $\mathbf{x}_{\mathrm{h}}$ and $1,000$ current data $\mathbf{x}_{\mathrm{c}}$ from $\mathrm{N}(\mathbf{0}, \boldsymbol{\Sigma}_{\mathbf{x}_{\mathrm{h}}})$ and $\mathrm{N}(\mathbf{0}, \boldsymbol{\Sigma}_{\mathbf{x}_{\mathrm{c}}})$, respectively.

We use two graphs as the graph of historical data $\mathcal{G}_{\mathrm{h}}$:
\begin{itemize}
    \item Erd\H{o}s--R\'enyi (ER) graph with edge connection probability $p = 0.15$.
    \item Regular sensor (RS) graph where each node is connected to the nearest eight neighbors.
\end{itemize}
These graphs have $N=100$, the edge weights of the ER graph are selected randomly from the uniform distribution $\mathrm{U}(1, 3)$, and those of the RS graph are given by $\mathbf{W}_{i,j} = \exp(\frac{ - \mathrm{dist}(i, j)}{\theta})$, where $\mathrm{dist}(i, j)$ is the Euclidean distance between nodes $i$ and $j$, and $\theta$ is a parameter.

We construct the current graph $\mathcal{G}_{\mathrm{c}}$ from $\mathcal{G}_{\mathrm{h}}$ with two types of topology changes: edge changes and node changes.
In the case of edge changes, we randomly remove $e$ edges from $\mathcal{E}_{h}$ and add $e$ edges connecting two randomly selected nodes.
In the case of node change, we randomly remove $v$ nodes from $\mathcal{V}_{h}$ and add $v$ nodes connected to other nodes with connection probability $p_{v} = 0.15$.

We compare the performance of the following methods: the estimator with ARMA graph filter computed from \eqref{eq:simple_gftr} (heareafter called ARMAE), and the estimator with ARMA graph filter computed by the proposed method based on density ratio weighting (ARMAE-DRW).
In this experiment, we perform $10,000$ Monte Carlo simulations and calculate the average MSE.

The results under edge changes and node changes are summarized in Tables \ref{tab:echange} and \ref{tab:nchange}.
As can be seen in these tables, ARMAE-DRW outperforms ARMAE for all datasets.
Fig.~\ref{fig:vis_gsr} shows the visualizations of the graph signal recovery results on the RS graph.
This figure demonstrates that ARMAE-DRW can recover the graph signal better than ARMAE.

\section{Conclusion}
In this paper, we proposed a graph filter transfer method, which learns a parametric graph filter in the situation that the graph of current data is different from that of historical data.
The proposed method estimates the probability density ratio of historical and current data and leverages it to learn a parametric graph filter that minimizes the MSE for current data.
The experimental results on missing value interpolation of graph signals demonstrated that the estimator constructed by the proposed method can successfully recover graph signals.

\bibliographystyle{IEEEbib}

\end{document}